\documentclass[preprint]{aastex}

\def\pix{pix$^{-1}$}
\def\deg{$^\circ$}

\def\mic{{$\mu$m}}

\def\h2o{H$_2$O}

\def\teff{$T_{\rm eff}$}
\def\mbol{$M_{\rm bol}$}

\def\aple{$\mathrel{\hbox{\rlap{\hbox{\lower4pt\hbox{$\sim$}}}\hbox{$<$}}}$ }
\def\apge{$\mathrel{\hbox{\rlap{\hbox{\lower4pt\hbox{$\sim$}}}\hbox{$>$}}}$ }

\begin{document}

\title{The Stellar Content of Obscured Galactic Giant H~II Regions: II. W42}

\author{R. D. Blum\altaffilmark{1}} \affil{Cerro Tololo Interamerican
Observatory, Casilla 603, La Serena, Chile\\ rblum@noao.edu}

\author{P. S. Conti} \affil{JILA, University of Colorado\\Campus Box
440, Boulder, CO, 80309\\pconti@jila.colorado.edu}

\author{A. Damineli}
\affil{ IAG-USP, Av. Miguel Stefano 4200, 04301-904, S\~{a}o Paulo, Brazil
\\damineli@iagusp.usp.br}

\altaffiltext{1}{Visiting Astronomer, Cerro Tololo Interamerican
Observatory, National Optical Astronomy Observatories, which is
operated by Associated Universities for Research in Astronomy, Inc.,
under cooperative agreement with the National Science Foundation}

\begin{abstract}

We present near infrared $J$, $H$, and $K$ images and $K-$band spectroscopy
in the giant H~II region W42. A massive star cluster is revealed; 
the color--color plot and $K-$band 
spectroscopic morphology of two of the brighter objects suggest the
presence of young stellar objects. The spectrum of
the bright central star is similar to unobscured stars with MK 
spectral types of O5--O6.5. If this star is on the zero age main 
sequence, then the derived spectrophotometric distance
is considerably smaller than previous estimates. The 
Lyman continuum luminosity of the cluster is a few times that of the
Trapezium. The slope of the $K-$band luminosity function is similar to
that for the Trapezium cluster and significantly steeper than that for
the massive star cluster in M17 or the Arches cluster near the Galactic center. 

\end{abstract}


\section{INTRODUCTION}

In this paper we continue the exploration of the stellar content of 
obscured Galactic
giant H~II regions begun by \citet[ hereafter Paper I]{bdc99}. 
$J$, $H$, and $K$ images are used to make a broad assessment of the stellar
content of obscured star forming regions in the Milky Way. Infrared spectroscopy
follows, providing details of the brightest cluster members which can be
used to make distance, mass, and luminosity estimates. 
The spectra
are placed in proper context by comparison to new infrared spectral 
classification systems for massive stars \citep{hcr96, brcfs97, fmn97,
hrl98}. The known hot star content of the Galaxy is rapidly expanding 
on the strength of the sophisticated infrared detector gains of the last
10 years. \cite{hhc97} have recently published a very detailed account of
the most massive stars in the (relatively) nearby giant H~II region M17. 
The present project (Paper I, this work,
and future work) seeks to provide a
large sample of massive star clusters 
with which to study the young and massive stellar content
in the Galaxy.  This sample builds on the detailed visual studies
of the Galactic OB associations \citep{mjd95} and 
provides a bridge to studies of young stellar objects in star forming regions.
Finally, the investigation of a large sample of clusters 
in Galactic giant H~II regions will be important in understanding the 
massive star clusters in the Galactic center \citep{ceal96, fmm99, feal99} 
which may have formed under different conditions than are
typical in the disk of our Galaxy \citep{ms96}.

W42 is located in the fourth Galactic quadrant at {\it l,b} $=$ 25.4\deg, 
0.2\deg. \citet{ldw85} determined W42 to be at the ``near'' kinematic distance
(3.7 kpc for $R_{\circ} =$ 8 kpc) and hence somewhat less luminous than 
earlier estimates \citep{sbm78}. 
In this series of papers we shall follow a suggestion of Dr.
Robert Kennicutt (private communication) that ``giant'' means that more
than $10^{50}$ Lyman continuum ($=$ Lyc) photons are inferred to be emitted per
second from the H~II region. This is about ten times the luminosity of the
Orion nebula and roughly the number emitted from the hottest {\it single}
O3-type star.  As these stars are not found in isolation, there is an
implication that a ``giant'' H~II region contains some minimum of {\it
multiple} O-type stars. In light of our new distance estimate (see \S4), W42
probably falls below this limit being perhaps a few times more 
luminous than the Trapezium in Orion. Our target list
was originally based on 
eleven of the most luminous giant H~II regions from the 
study of \citep{sbm78} who tabulated their Lyc output derived from radio 
continuum measurements and kinematic distance estimates.
W42 is not seen in visual images,
and we estimate a foreground extinction of $A_V$ $=$ 10 mag (see below).

\section{OBSERVATIONS AND DATA REDUCTION}

$J$ ($\lambda \sim 1.3$ \mic, $\Delta\lambda \sim 0.3$ \mic), 
$H$ ($\lambda \sim 1.6$ \mic, $\Delta\lambda \sim 0.3$ \mic), and 
$K$ ($\lambda \sim 2.2$ \mic, $\Delta\lambda \sim 0.4$ \mic)
images of W42 were obtained on the nights of 29
August 1998 and 01, 02 May 1999 with the f/14 tip--tilt system on the
Cerro Tololo Interamerican Observatory (CTIO) 4m Blanco telescope
using the two facility imagers CIRIM (1998 data) and OSIRIS, the Ohio State
InfraRed Imager/Spectrometer\footnote{OSIRIS is a collaborative
project between the Ohio State University and CTIO. OSIRIS was
developed through NSF grants AST 9016112 and AST 9218449.}
(1999 data).  Spectroscopic data were obtained on the night of
02 May 1999 using the f/14 tip--tilt system at the Blanco telescope
with OSIRIS. CIRIM and OSIRIS are described in the instrument manuals
found on the CTIO web pages (www.ctio.noao.edu). For OSIRIS, see also
\citet[]{dabfo93}. The tip-tilt system is
described by \citet[]{pe98}. The
tip--tilt system uses three piezo--electric actuators to move the
secondary mirror at high frequency in a computer controlled feed--back
loop which corrects the natural image centroid motion. OSIRIS employs 0.16$''$
pixels and CIRIM 0.21$''$ pixels.

All basic data reduction was accomplished using IRAF\footnote{IRAF is
distributed by the National Optical Astronomy Observatories.}. Each
image/spectrum was flat--fielded using dome flats and then sky
subtracted using a median combined image of five to six frames. For
W42 itself, independent sky frames were obtained five to ten arcminutes 
south of the cluster. Standard stars used the median
combination of the data for sky.

\subsection{Images}

The OSIRIS 1999 May images were obtained under photometric conditions
and in $\sim$ 0.5$''$ to 0.6$''$ FWHM seeing (with the tip--tilt
correction). Total exposure times were 270~s, 135~s, and 135~s at $J$,
$H$, and $K$, respectively.  The individual $J$, $H$, and $K$ frames
were shifted and combined (Figure~\ref{w42i}), and these combined frames
have point sources with FWHM of $\sim$ 0.6$''$, 0.7$''$, and 0.6$''$ at 
$J$, $H$, and $K$, respectively. DoPHOT \citep[]{sms93} photometry was 
performed on the combined
images. The flux calibration was accomplished using standards 9170 and
9172 from \citet[]{peal98} which are on the Las Campanas 
Observatory standard system (LCO). The LCO standards are essentially on the
CIT/CTIO system \citep[]{efmn82}, though color 
transformations exist between the two systems for redder stars.
The standards were taken just before the W42 data and within 0.16 airmass
of the airmass for W42; no corrections were applied for these small
differences in airmass. Aperture corrections using 11 pixel radius
apertures were used to put the instrumental magnitudes on a flux
scale. 

Stars brighter than about 10th magnitude are expected to be a few
percent non-linear on the OSIRIS images. We have included the 1998 CIRIM 
data for such stars
since the count levels for the CIRIM images were in the fully linear
regime (no linearity correction needed). 
The zero point to the CIRIM photometry was determined by 
comparing the instrumental magnitudes of stars 
in common to the OSIRIS and CIRIM images.
The CIRIM $J-$band photometry includes a color correction term 
in order to make a transformation on to the CIT/CTIO system (PAPER I).
The OSIRIS data
was placed on the CIT/CTIO system by making small corrections ($<$ 10 $\%$)
to the $J$, $H$ and $K$ magnitudes based on linear fits to the 
magnitude differences as a function of color for 
stars in common to the CIRIM and OSIRIS images.

Uncertainties for the final $JHK$ magnitudes include the formal DoPHOT
error added in quadrature to the error in the mean of the photometric
standards (including the transformation to OSIRIS magnitudes for the
CIRIM data), and the error in aperture corrections used in
transforming from the DoPHOT photometry to OSIRIS magnitudes.  The
latter errors dominate and were derived from the scatter in the
measurements of four to seven relatively uncrowded stars on the mosaic
frames. The sum (in quadrature) of the aperture correction and
standard star uncertainties is $\pm$ 0.018, $\pm$ 0.023, $\pm$ 0.019
mag in $J$, $H$, and $K$, respectively. The DoPHOT errors ranged from
approximately $\pm$ 0.01 mag to an arbitrary cut--off of 0.2 mag
(stars with larger errors were excluded from further analysis).

The flat--field illumination was not uniform. A smooth gradient with full 
range of about 10$\%$ was present. Corrections for this gradient 
were made based on observations of
a standard star taken over a 49 position grid covering the array.

\subsection{Spectra}

The spectra of three of the brightest four stars in the 
center of W42 were obtained with a 0.48$''$ wide slit (oriented EW) in \aple
0.7$''$ FWHM seeing  and divided by the spectrum of HR~6813 (A1V) to remove
telluric absorption features. Br$\gamma$ absorption in HR~6813  was removed
by eye by drawing a line across it between two continuum points. One
dimensional spectra were obtained by extracting and summing the flux in a
$\pm$ 2 pixel aperture (0.64$''$ wide).
The extractions
include background
subtraction from apertures centered \aple 1.0$''$ on either side of the object.

The wavelength calibration was accomplished by measuring the positions of
bright OH$^-$ lines from the $K-$band sky spectrum 
\citep[]{oo92}. Lines are
identified by their relative differences between one and another. The
measured dispersion is 0.0003683 \mic \ pix$^{-1}$. The spectral resolution
at 2.2 \mic \ is $\lambda/\Delta\lambda \approx$ 3000. 

\section{RESULTS}

A spectacular stellar cluster is revealed at the heart of W42 
in our near infrared images (Figure~\ref{w42i} and 
Figure~\ref{w42c}). Apparently, the cluster 
has just emerged from
the edge of the molecular cloud from which
it formed. This is confirmed below through $K-$band spectroscopy which
shows that the central massive star has largely cleared away its birth 
cocoon, but two of the next brightest stars have not.
\citet{ldw85} observed W42 in the mid and far infrared. Their analysis 
revealed two distinct sites of star formation toward W42. 
One, G25.4SE is located at 3.7 kpc from the sun (accounting for a
sun to Galactic center distance of 8 kpc), 
while the other, G25.4NW is probably located
at about 9.6 kpc. Lester et al. found a 10 \mic \ unresolved source located at
RA (2000) $= 18^{\rm h}~38^{\rm m}~15^{\rm s}.2$, 
Dec (2000) $= -06$\deg$~47'~50''$ which is
coincident with G25.4SE. This mid IR source 
is associated with the stellar cluster we have observed. 
Comparison of the position
of the bright foreground star in the SE corner of
our $K-$band image to the same star on the image from
the Digitized Sky Survey (DSS)
\footnote{
Based on photographic data obtained using The UK Schmidt Telescope.     
The UK Schmidt Telescope was operated by the Royal Observatory          
Edinburgh, with funding from the UK Science and Engineering Research    
Council, until 1988 June, and thereafter by the Anglo-Australian        
Observatory.  Original plate material is copyright (c) the Royal        
Observatory Edinburgh and the Anglo-Australian Observatory.  The        
plates were processed into the present compressed digital form with     
their permission. The Digitized Sky Survey was produced at the Space   
Telescope Science Institute under US Government grant NAG W-2166.} 
results in a position for the ($K-$band) bright central star in the cluster 
(W42~\#1; see \S3.2) of 
RA (2000) $= 18^{\rm h}~38^{\rm m}~15^{\rm s}.3$, 
Dec (2000) $= -06$\deg$~47'~58''$.

\subsection{Images}

The $H-K$ color$-$magnitude diagram (CMD) for the region toward W42 is shown in
Figure~\ref{cmd}. A cluster sequence is evident at 0.6 \aple $H-K$ \aple 1.5
along with stars with much redder colors and a probable foreground sequence. 
The $J-H$ vs. $H-K$ 
color--color diagram is presented in Figure~\ref{cc}. As expected from the 
morphology and range of colors in Figure~\ref{w42c}, 
the strong effects of differential reddening can 
be seen in the color--color diagram. Typical reddening lines are shown for 
M giants \citep{fpam78} and early O stars \citep{k83}. The latter
color was transformed as described in \S3.1.3.
The cluster stars defined below in \S3.1.1 and
shown as {\it open circles}
in Figure~\ref{cc} may have a slight unexplained 
systematic offset relative to the 
reddening line for normal O stars. This possible offset will not affect the
conclusions in this paper regarding the cluster stars.
The relationship for the intrinsic colors of
classical T~Tauri stars (pre--main sequence stars) \citep{mch97} is also shown
for reference.
The adopted reddening law
is from \citet{m90}. 

\subsubsection{The Central Cluster}

By separating the stars in Figure~\ref{cmd} based on radial position, we can
better define the central cluster relative to the surrounding field. In 
Figure~\ref{surf}, we plot the radial surface density of stars centered on the
position of the bright central star, W42~\#1 (\S3.2). 
We also plot the radial surface
density for stars with $K$ $\leq$ 14 mag, for which the number counts are 
more nearly complete (see below).
The surface density
becomes approximately uniform at a radius of $R \sim 30''$, continuing up
to 50$''$. It then begins to 
fall rapidly at the edge of the array ({\it dashed vertical line}) as expected
due to the rectangular shape of the field. 
Taking these radii as representative, 
we divide the
CMD into regions with 0$'' \leq R <$ 30$''$, 30$'' \leq R <$ 50$''$, and 
$R >$ 50$''$,
as shown in Figure~\ref{cmdrad}. These represent the central cluster, 
a background annulus, and the region of the array where edge effects are 
important to the radial number counts.

The range of color and brightness in Figure~\ref{cmdrad} overlaps for stars
in the cluster and background regions.
A distinct blue sequence can be seen in the 
0$'' \leq R <$ 30$''$ cluster CMD which appears to merge smoothly 
with other stars in the 30$'' \leq R <$ 50$''$ CMD. Clearly the 
cluster can not be completely
extracted from the surrounding field based on radial position alone.
In order to further enhance the cluster sequence, we defined 
a background CMD using the 30$'' \leq R <$ 50$''$ region and accounting for
the area difference between this annulus and the central region. 
The background CMD was binned in 0.5 mag color--magnitude bins.
We then randomly selected and subtracted stars from the cluster in 
equal numbers from bins matching the background CMD. 
The resulting CMD is shown in Figure~\ref{cmdsub}. The corresponding colors
for these stars are plotted as {\it open circles} in Figure~\ref{cc}.

\subsubsection{Extinction to the Cluster}

Most of the brightest stars in the central cluster
fall along an essentially vertical track as expected for hot stars. The average
color of the brightest seven of these stars
is $H-K$ $=$ 0.637 which corresponds to
an extinction at 2.2 \mic \ 
of $A_K =$ 1.07 mag ($A_V$ $\approx$ 10 mag) using the 
interstellar reddening curve of \citet{m90} and an intrinsic $H-K$ $=$ $-0.04$
\citep{k83}. 
Many other stars appear more reddened, typically 
up to $H-K$ $\sim$ 2 with
some as high as 3.8. 
A star with $H-K$ $=$ 2 and intrinsic $H-K$ $=$ 0 would
have  $A_K =$ 3.2 mag ($A_V$ $\approx$ 32 mag).

\subsubsection{$K$ vs. $H-K$ CMD for the ZAMS}

It will be useful in discussing the W42 cluster below 
to have an estimate for the 
zero age main sequence (ZAMS) transformed into the $K$ vs. $H-K$ plane. We
have constructed such an estimate using the model results of \citet{ssmm92}.
The models give bolometric luminosities (\mbol) 
and effective temperatures (\teff)
for stars of a given mass as they begin their evolution on the ZAMS. 
Using relationships for spectral type vs. \teff \ \citep{vgs96, j66}, 
visual bolometric correction (BC$_V$) vs. effective temperature \citep{vgs96,
mmrk86}, $V-K$ vs. spectral type \citep{k83}, and $H-K$ vs. spectral type 
\citep{k83}, we transform the \mbol \ and \teff \ to the $K$ vs. $H-K$ CMD.
A small correction has been made in an attempt to place the $H-K$ colors from
\citet{k83} onto the CIT/CTIO system. The \citet{k83} colors are basically 
referred to the
system of \citet{j66} with newly defined colors involving $H$ (Johnson
used no $H$ filter). In addition, \citet{k83.1} found that interpolating an
$H$ magnitude from the observed 
$J$ and $K$ values gives a very good estimate
of the observed $H$ magnitude.
\citet{c90} found no discernible color correction between
the SAAO system and that of \citet{j66}. We have therefore
used the transformation
of \citet{c90} between the SAAO system and CIT/CTIO to transform 
the $H-K$ colors given by \citet{k83}. These corrections are
at most one percent and hence negligible compared to the 
measurement uncertainties and scatter due to differential reddening.

The ZAMS CMD is shown in Figure~\ref{cmdsub} for a particular distance (2.2 kpc)
which is discussed in \S4. The values plotted in Figure~\ref{cmdsub} are listed
in Table~\ref{tabzams}. These values may be compared to those given by 
\citet{hhc97}. There is generally good agreement as a function of \teff. This
is expected since the same models and colors were used in both cases. There are 
systematic 
differences at the $\sim$ $\pm$ 0.3 mag level for $M_K$ as a function of 
spectral type since the 
spectral type vs. \teff \ relation adopted here \citep{vgs96, j66} 
is different than
that used by \citet{hhc97} who used the relation given by 
\citet{mpg89}. These differences do not affect the
conclusions of \S4.3 since the resulting $M_K$ for the spectral type in question
(O5--O6) is within 0.1 mag in both cases.

\subsubsection{The $K-$band Luminosity Function}

The $K-$band luminosity function (KLF) is shown in Figure~\ref{klf}.
Neglecting the last five bins where the counts appear to be incomplete and
the first 3 bins where the counts may be better described
as uniform, the $K-$band counts are
well fit by a power--law with index 0.4 (log$_{10}$N $=$ 0.4 $\times$ $K$ 
+ constant). 

We estimated the completeness of the KLF by performing artificial star experiments. Initially, we attempted to add stars to the original $K-$band image, but 
even adding 10$\%$ of the observed KLF resulted in recovered luminosity functions much less complete than the observed one. This is due to the high spatial density of the cluster. To avoid this problem, we constructed complete artificial
frames and analyzed them in the same way as the original frame. 
Anticipating the result, we used the spatial density distribution 
of the stars on the $K-$band image with $K <$ 14 mag (Figure~\ref{surf})
to generate the positions of stars on the artificial frames. The
input luminosity function was constructed from two components, a uniform 
distribution from 8.5 $\leq K \leq$ 10.5, and a power--law for 10.5 $< K \leq$ 19.5. The total number of stars was set by the flux on the $K-$band
image, less a uniform background component determined from the sky frame. 
This should be conservative regarding the number of stars
since the strong nebular flux is included. This latter aspect may be 
balanced somewhat by the fact that the artificial frames did
not include variable extinction and nebular emission.
The total $K-$band flux in Figure~\ref{w42i}$c$ is 6.0 mag, 
not including the bright saturated star in the SE corner of the frame.
The total flux of the average actual 
input luminosity function (Figure~\ref{fake})
was 5.9 mag which can be compared to the total flux in observed stars
(Figure~\ref{klf}), $K =$ 6.18 mag. 

We constructed 10 artificial frames by randomly sampling the spatial 
distribution and luminosity functions. DoPHOT was run on each frame and the 
recovered stars matched to the input lists. The average recovered luminosity
function, input luminosity function, and completeness fraction are shown
in Figure~\ref{fake}. The general shape and distribution of the recovered
stars in Figure~\ref{fake} suggest the experiments are a fair test of 
completeness of the original frame. From the same input and recovered star 
lists, we also show the luminosity functions and completeness fraction
for the stars located at 0$'' \leq R <$30$''$. See Figure~\ref{fake30}.

Tests were also made to check whether somewhat steeper power--laws were also
consistent with the observed KLF. This might be the case if the crowding
were sufficient to ``hide'' many fainter stars. Artificial star experiments
analogous to those above (including the same uniform component) 
but with a power--law component of 0.5 are not 
consistent with the data. A single power--law component of this steepness
produces more star light than is necessary to account for the observed total
flux and  
produces too many stars in the recovered luminosity function at fainter
magnitudes if it is required to produce approximately the correct numbers
at brighter magnitudes.

\subsection{Spectra}

The spectra of the targets W42~\#1, \#2, and \#3 are shown in Figure~\ref{spec}.
The final signal--to--noise in the three spectra is typically 80--105, 75--95,
and 55--78 for W42~\#1, W42~\#2, and W42~\#3, respectively and is higher on the
red end than the blue end for all three.
The brightest star in the central cluster, W42~\#1 ($K = $ 8.8 mag), shows
characteristic O star features \citep{hcr96}. These include \ion{C}{4}
(2.069 \mic \ and 2.078 \mic) emission, \ion{N}{3} (2.1155 \mic) emission, Br$\gamma$ (2.1661
\mic) absorption, and \ion{He}{2} (2.1891 \mic) absorption.
Comparison to the standards
presented by \citet{hcr96} results in $K-$band spectral type of kO5-O6. 
These stars typically have MK spectral types of O5 to O6. The present 
classification system laid out by \citet{hcr96} does not have strong luminosity 
class indicators. Still, the \ion{He}{1} (2.06 \mic) and Br$\gamma$ features 
can be used to approximately 
distinguish between dwarf+giants on the one hand, and supergiants on
the other: the supergiants tend to have emission or weak absorption 
in these lines. The
spectrum of W42~\#1 shown in Figure~\ref{spec} has been background subtracted
with nearby apertures to account for the nebular emission seen in projection 
toward the star. The apparent absorption feature at the position of Br$\gamma$
and absence of a feature at \ion{He}{1} (2.06 \mic) (which might indicate 
a poor subtraction of the nebular contribution), suggests that 
W42~\#1 is a dwarf or giant star.

In contrast to W42~\#1, the spectra of W42~\#2 and \#3 show only emission 
features at \ion{He}{1} (2.06 \mic) and Br$\gamma$. 
These spectra have also been 
background subtracted with nearby (\aple 1.0$''$) background 
apertures. We believe the strong emission remaining after this subtraction
is related to the 
local environment in these stars (see \S4.1).

\section{DISCUSSION}

\begin{figure}
\epsscale{0.4}
\plotone{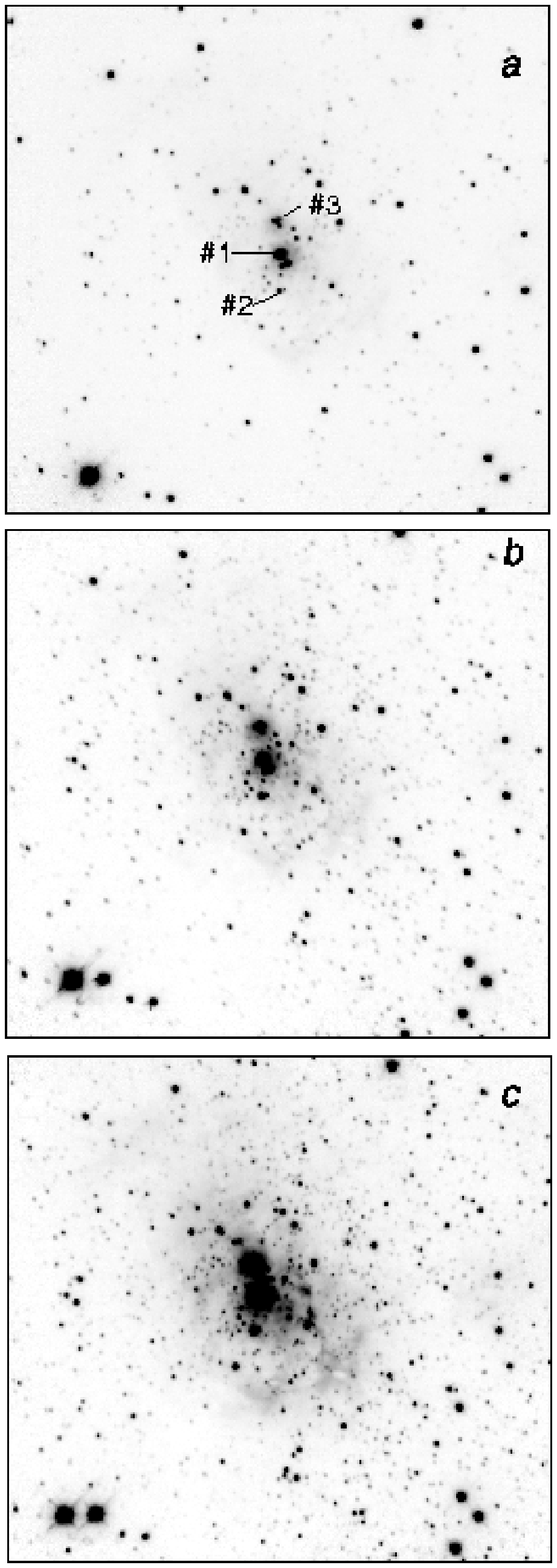}
\figcaption[]{$JHK$ images of the massive star cluster in W42.
North is up, East to the left, and the scale is 0.16$''$ pix$^{-1}$ 
in these $\sim$
1.8$'$ $\times$ 1.7$'$ images. The combined images have point sources 
with FWHM of approximately 0.6$''$, 0.7$''$, and 0.6$''$ at $J$, $H$, and $K$, 
respectively.
$a$) $J-$band.
$b$) $H-$band. $c$) $K-$band.
Stars with spectra presented in Figure~\ref{spec} are labeled in $a$. W42~\#3
is not the most prominent source at $J$ (but is at $K$) 
near the line marking its position.
See also Figure~\ref{w42c}.
\label{w42i}
}
\end{figure}

\begin{figure}
\epsscale{0.8}
\plotone{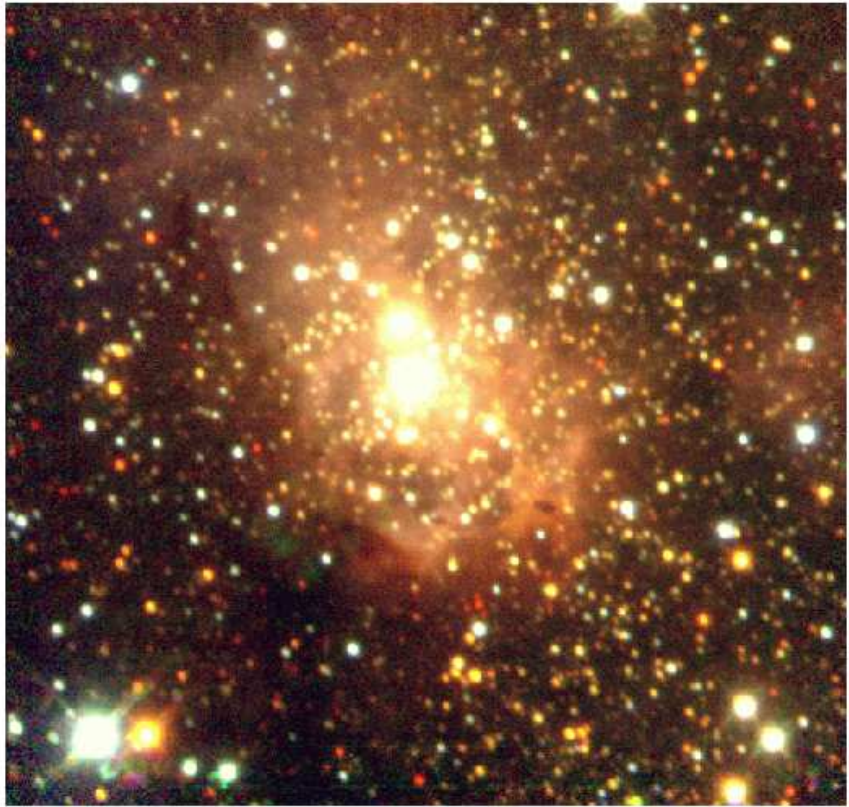}
\figcaption[]{
False color image ($K =$ red, $H =$ green, $J =$ blue) constructed from the
images shown in Figure~\ref{w42i}.
North is up, East to the left.
\label{w42c}
}
\end{figure}

\begin{figure}
\plotone{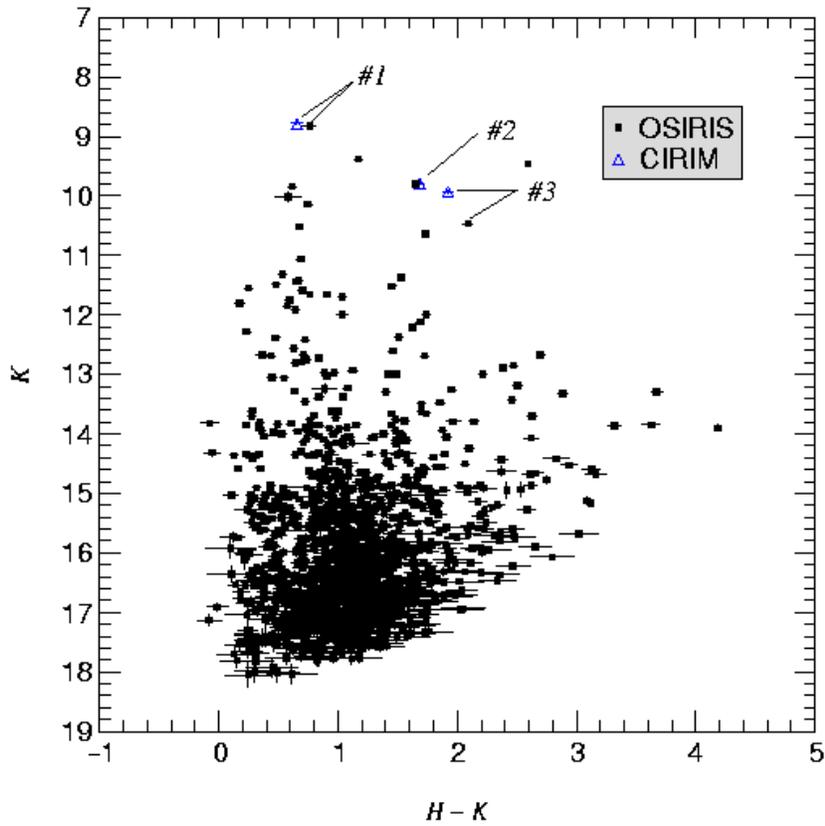} 
\figcaption[]{$H-K$ color--magnitude diagram (CMD)
for the W42 cluster and surrounding field. Stars for which spectra are 
presented are labeled. The {\it open triangles} are from the CIRIM data and are 
plotted for stars brighter than the OSIRIS saturation limit where
possible (the CIRIM field of view is smaller than that for OSIRIS).
\label{cmd}
}
\end{figure}

The dense stellar cluster evident in Figure~\ref{w42i} surrounded by 
intense nebulosity leaves no doubt that this is a young object still 
emerging from its birth environment at the edge of its parent molecular cloud. 
The cluster appears to have emerged by clearing the foreground
material to the West; darker regions with fewer stars remain toward the East.
There is a suggestion of photoevaporated regions on the edge of the 
cloud to the East similar to those
seen in M16 \citep{heal96} and NGC~3603 \citep{beal99}. 
This is particularly
evident in Figure~\ref{w42i}$a$. 
This picture is consistent with the CO line maps of \citet{ldw85}, who found
the peak brightness temperature of the associated molecular cloud to be 
offset to the East of the W42 (G25.4SE) H~II region (see their Figure~8).
Higher spatial resolution images and in 
nebular lines are in order to further study the interaction and impact of the 
ionizing cluster on the molecular cloud interface. 

The process of clearing the local environment
is still on--going, clumps of dark material are seen in projection against the
photoionized H~II region (e.g. SW of the cluster center).
In \S3.1.2, the extinction to the brightest stars on the cluster sequence
was found to be $A_K$ $=$ 1.07 mag. 
This value can be taken as representative of the foreground
extinction to the cluster, and indicates a $V$ magnitude of approximately
18 for W42~\#1. This is consistent with the non--detection of the cluster
on the DSS plates. 
A few stars remain after subtracting the "background"
component from the CMD (\S3.1.1), but these are more likely in the foreground
since the extinction toward them is as low as $A_K$ $\sim$ 0.3 mag. 

\subsection{Young Stellar Objects}

There are a host of very red objects indicated in Figures~\ref{w42i}, 
\ref{cmdrad}, and
\ref{cmdsub}. Such colors are suggestive of young stellar objects (YSO) 
in the context of an embedded stellar cluster emerging from its parent cloud.  
The observed colors of known
populations of pre--main sequence stars (PMS) and more heavily 
embedded protostars appear to lie on/occupy 
rather well defined sequences/regions in observational color space;
and their
colors, which are redder than usual stellar photospheres have been 
successfully modeled as arising from excess emission produced by circumstellar
disks \citep{la92, mch97} and envelopes \citep{hkc93, psl97}. 
Like normal stars, PMS objects may be seen
through foreground extinction, further reddening their colors. Deeply embedded
protostars may have colors with or without excess emission, but typically 
suffer large extinction due to the dense envelope surrounding them 
\citep{la92}. Obviously, protostars too may suffer additional foreground
extinction.

\begin{figure}
\plotone{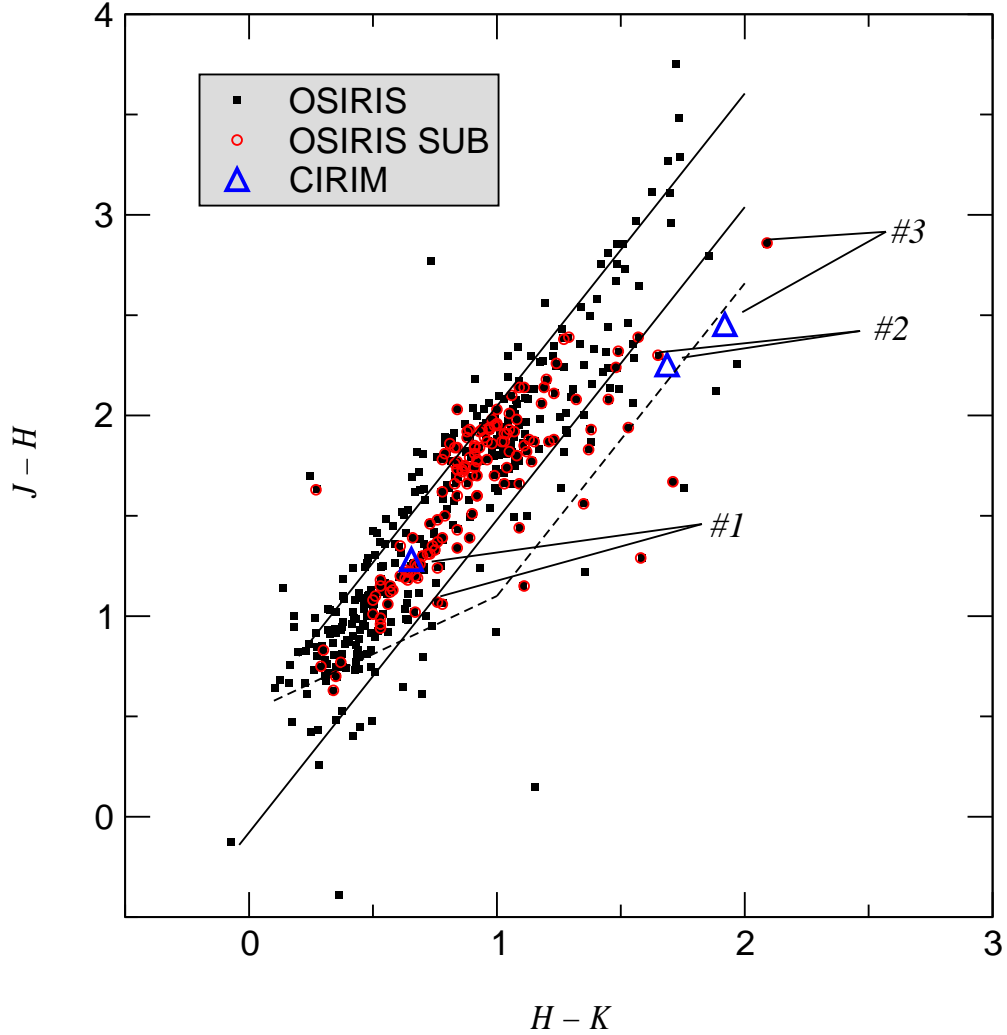}
\figcaption[]{$J-H$ vs. $H-K$ color--color diagram for the the W42 cluster.  
All stars in the field detected at $J$, $H$, and $K$ are plotted as
{\it filled black squares}. Stars in the final statistically subtracted
central cluster are plotted as {\it open circles}.
The CIRIM data for 
W42~\#1, \#2, and \#3 are also
presented ({\it open triangles}). 
For comparison, the reddening lines for early O stars (lower line, $H-K$ $=$
$-$0.04, $J-H$ $=$ $-$0.14, \citep{k83}; see text) 
and M giants (upper line, $H-K$ $=$ 0.2, $J-H$ $=$
0.8, \citep{fpam78}) are shown.  The classical T~Tauri star locus \citep{mch97}
is also shown for reference ({\it dashed line}).
W42~\#3 appears to have varied significantly in both colors; see text. 
\label{cc}
}
\end{figure}

\begin{figure}
\plotone{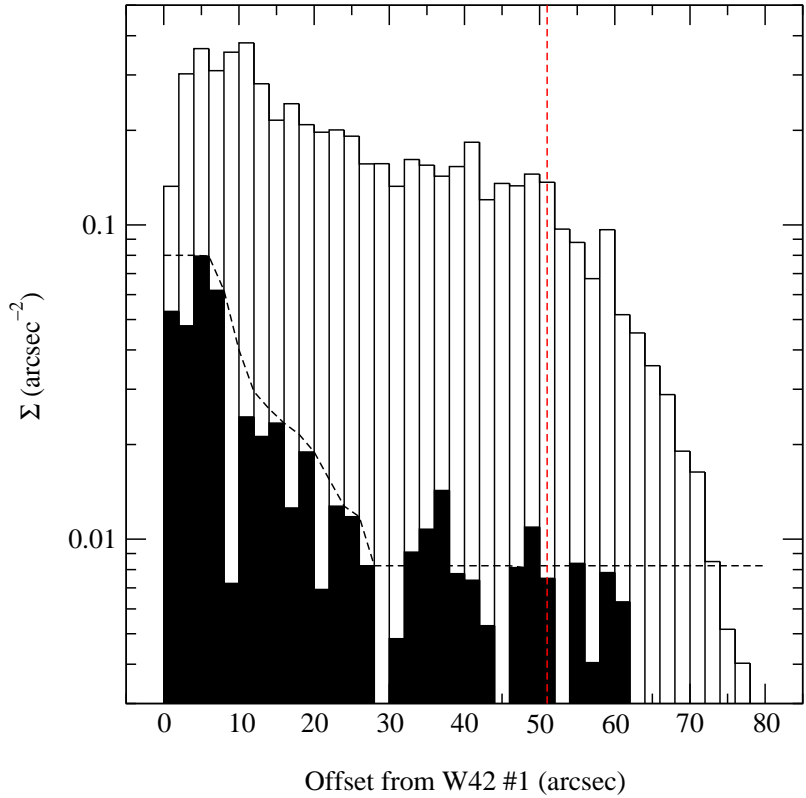}
\figcaption[]{
The observed $K-$band radial surface density. 
Radius is defined as distance in arcsec from the bright central star
W42~\#1.
A central cluster can be defined
at $R <$ 30$''$ where the counts become uniform. The {\it dashed vertical
line} represents the array edge where the radial counts begin to fall as 
expected due to the rectangular shape of the field. The {\it solid} histogram
is for stars with $K$ $\leq$ 14 mag where the number counts are nearly
complete, and the curve represents the radial surface density
distribution used in 
the artificial star experiments to gauge completeness; see text.
\label{surf}
}
\end{figure}

\begin{figure}
\plotone{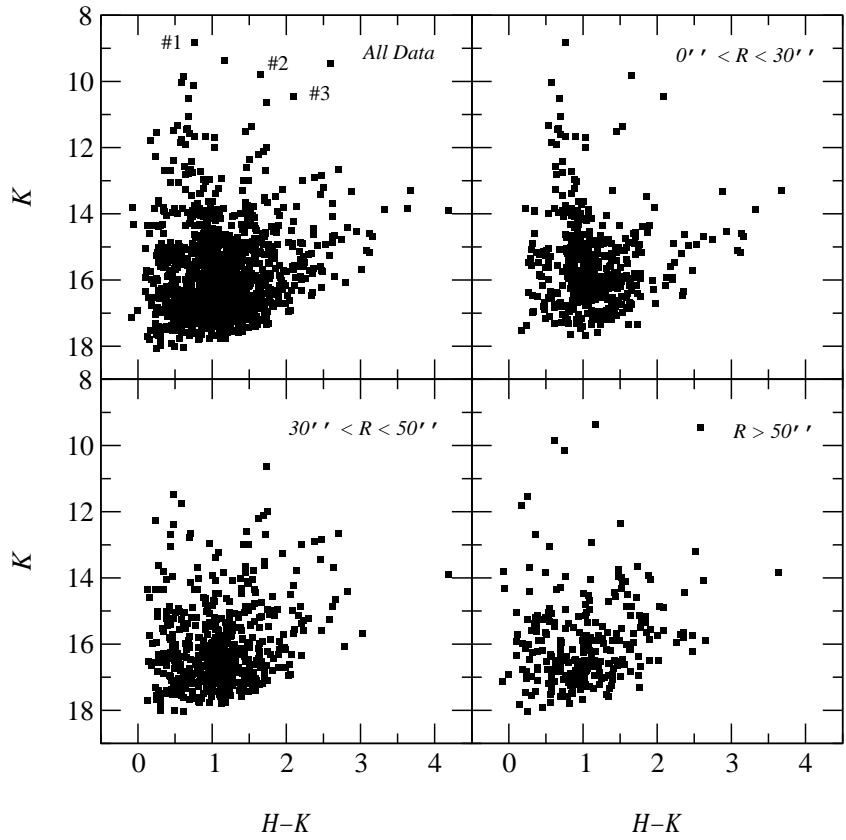}
\figcaption[]{
The $H-K$ color--magnitude diagram as a function of radial position. 
By selecting stars at $R <$ 30$''$ as indicated in Figure~\ref{surf}, 
the central cluster sequence becomes more well defined relative to the 
background.
\label{cmdrad}
}
\end{figure}

\begin{figure}
\plotone{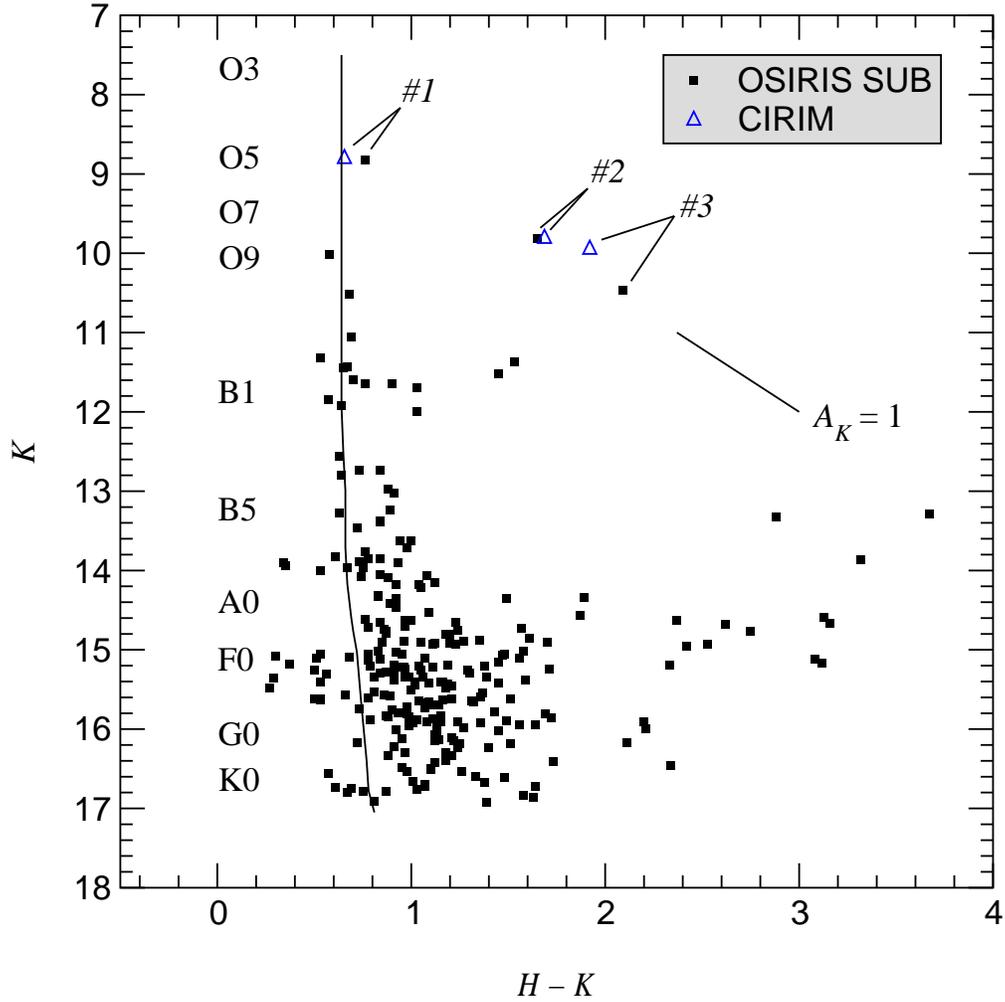}
\figcaption[]{
$H-K$ vs. $K$ color--magnitude diagram (CMD) for the central cluster. 
A background component has been subtracted based on the number and position 
of stars in the 30$''< R <$ 50$''$ CMD shown in Figure~\ref{cmdrad}; see text. 
The {\it solid line} shows the relationship for the ZAMS based on the 
Z$ =$ 0.02, high mass--loss models of
\citet{ssmm92}; see Table~\ref{tabzams}. 
The transformation to the observational plane is discussed in
the text. CIRIM data ({\it open trinagles}) are also shown for 
W42~\#1, W42~\#2, and W42~\#3; see Figure~\ref{cmd}.
\label{cmdsub}
}
\end{figure}

\begin{figure}
\plotone{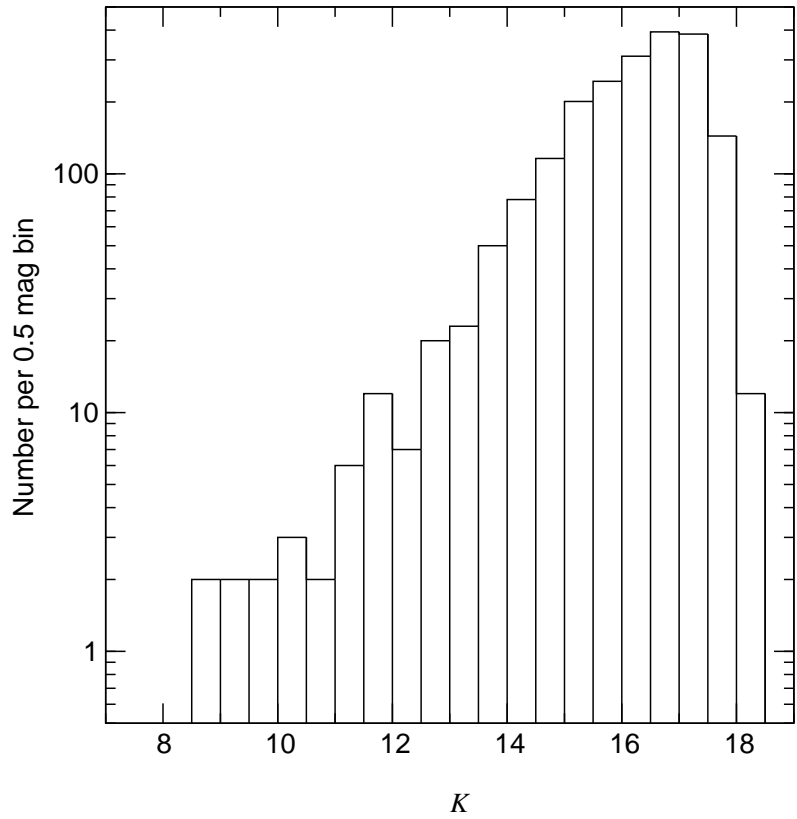}
\figcaption[]{
The observed $K-$band luminosity function (KLF) for the entire area shown in 
Figure~\ref{w42i}$c$. No corrections for extinction or completeness have been
made.
\label{klf}
}
\end{figure}

\begin{figure}
\plotone{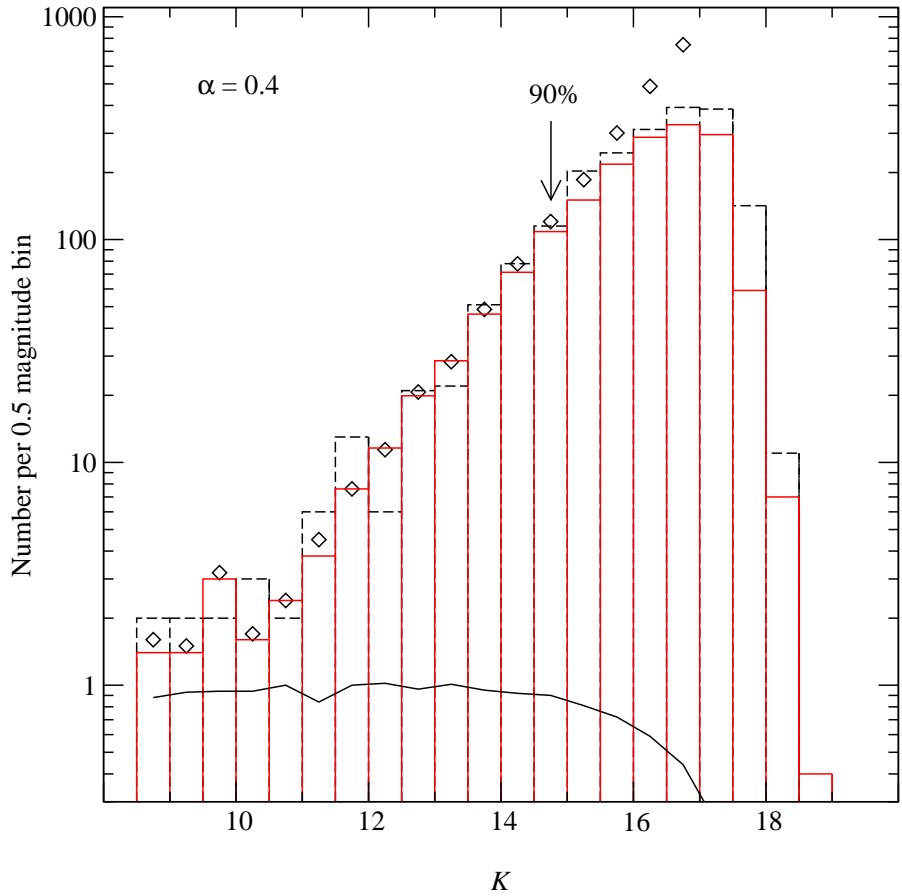}
\figcaption[]{
Artificial luminosity functions and the $K-$band completeness limit.
The input ({\it open diamonds}) luminosity function and recovered
({\it solid histogram}) 
luminosity function are shown for an average of 10 artificial frames. For
comparison, the observed $K$-band luminosity function is also plotted.
The input luminosity function consists of a uniform distribution 
for the brightest stars and a power--law with index 0.4 for the remainder; see
text for details. The ({\it solid line}) is a ratio of the recovered to 
input luminosity functions.
\label{fake}
}
\end{figure}

\begin{figure}
\plotone{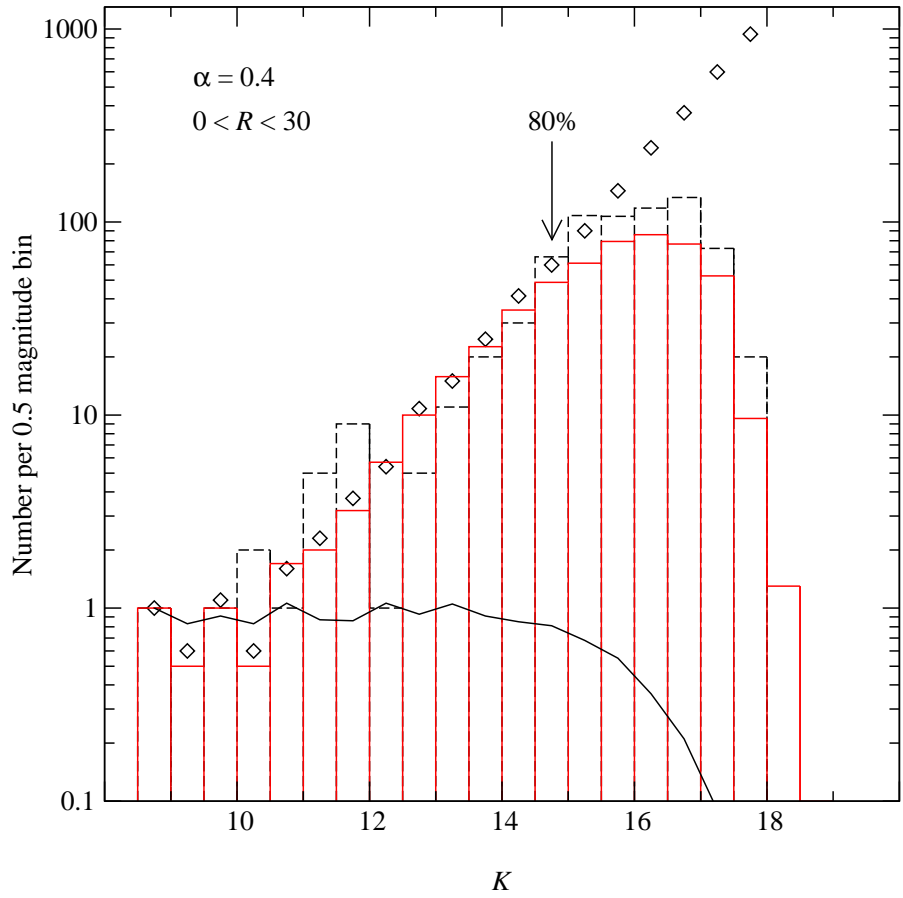}
\figcaption[]{
Same as Figure~\ref{fake} but for stars located in the central 30$''$ of the
cluster.
\label{fake30}
}
\end{figure}

\begin{figure}
\plotone{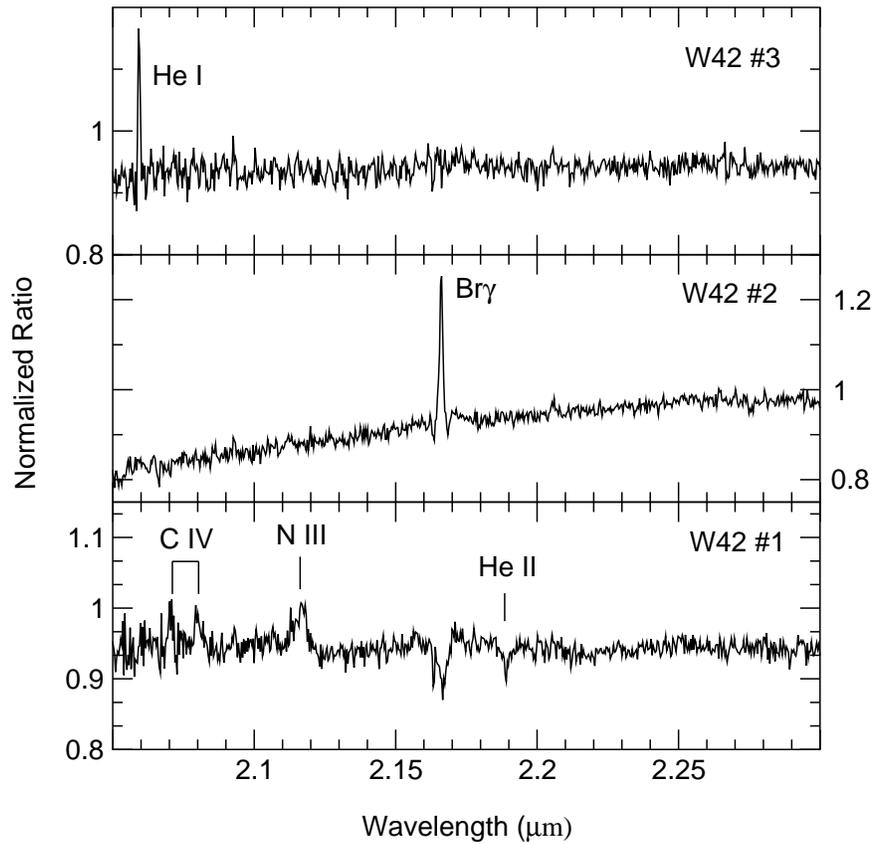}
\figcaption[]{
$K-$band spectra for three of the brightest stars in the 
W42 cluster.  The two pixel resolution is
$\lambda/\Delta\lambda \approx$ 3000. The spectra were summed in apertures
0.64$''$ wide $\times$ a slit width of 0.48$''$ and include background 
subtraction from apertures centered \aple 1.0$''$ on either side of the object.
\label{spec}
}
\end{figure}

The $J-H$ vs. $H-K$ 
color--color diagram is most useful in assessing whether any particular 
star may have excess emission. 
The diagram distinguishes between normal stellar colors which are seen through
a column of dust, hence making them redder, and a contribution which is
due to emission (reprocessed stellar light from
circumstellar envelope and/or accretion luminosity from a 
circumstellar disk). \citet{la92} and \citet{mch97} have shown that disks
can provide a source of excess emission to the normal stellar colors.
In Figure~\ref{cc}, we have plotted the classical T Tauri sequence (CTTS)
from \citet{mch97} along with reddening lines for the interstellar reddening
law of \citet{m90}. The former locus may be understood as arising from disk
luminosity and projection effects 
(e.g, Lada \& Adams 1992 and references therein).
The near infrared colors of weak lined T Tauri stars (WTTS)
are similar to normal stars \citep{mch97}. These stars exhibit
spectra which suggest they have disks, but the disk contribution to 
the near infrared colors is negligible. The Herbig AeBe stars which are higher
mass analogs to the CTTS show similar colors to the CTTS but 
extend to generally larger $H-K$ excess and with a larger 
contribution to the colors from circumstellar extinction \citep{la92}.
The colors of these objects have been fit by circumstellar envelope models in 
which the excess emission 
arises from reprocessed stellar light via dust heating and re-radiation
\citep{hkc93, psl97}, or alternatively, from circumstellar disks with central
holes \citep{la92}. 
We also show the reddening lines for a main sequence
O star and M giant (see S3.1) which can be taken as approximate guides for
the expected colors for normal stars along sight lines to the inner Galaxy.
The {\it black squares} in Figure~\ref{cc} represent all the stars in
Figure~\ref{w42i} for which $J$, $H$, and $K$ colors were measured. The
{\it open circles} represent the stars in the "background subtracted"
central cluster (\S3.1.1). 

The W42 cluster stars (including W42~\#1) 
occupy a rather tight sequence at modest
reddening in Figure~\ref{cc} as expected for young massive stars
seen through different columns of obscuring material. At larger 
reddening ($H-K$ , $J-H$ $=$ 1,2),
there is a larger dispersion of colors. There are also a number
of stars which lie significantly 
beyond the reddening band for normal stars in the
region of PMS stars and protostars as discussed above.
Two of these objects are W42~\#2 and \#3. Spectra for these objects are shown
in Figure~\ref{spec}. Neither spectrum shows stellar absorption features, though
both exhibit emission at Br$\gamma$ or \ion{He}{1} (2.06 \mic) after
background subtraction (\S2). Accounting for the foreground extinction to
W42~\#1, both W42~\#2 and \#3 lie in a region of the color--color plot 
occupied by luminous protostars and Herbig AeBe stars \citep{la92}. \citet{la92}
suggest that Herbig AeBe stars might be recently "uncovered" protostars given
the large overlap in the color--color plot for the two objects. 
We believe that the position of W42~\#2 and \#3 in the color--color plot
combined with their emission--line spectra,
strongly suggests they are also luminous YSOs. The absence of absorption
features is consistent with veiling due to 
the excess emission seen in Figure~\ref{cc}.
In the case of W42~\#2, the residual Br$\gamma$ emission may be due
to a circumstellar disk;
spherical or envelope distributions to the circumstellar material
are also possible. Higher spectral resolution data on the Br$\gamma$ 
line in this object will be required to
rule out models of one nature or the other.
In W42~\#3, the presence of \ion{He}{1} emission but no
Br$\gamma$ could be due to imperfect subtraction. W42~\#3 has closer neighbors 
making the subtraction of the background more difficult. If the emission seen in
projection toward W42~\#3 is due to a 
compact H~II region, then the \ion{He}{1} 
emission may be coming from a region closer to the star than the Br$\gamma$,
the latter naturally subtracting off better on these angular scales.
We plan to obtain $H-$band spectra of W42~\#2 and \#3 where the excess emission
should be less allowing for an improved picture of the nature of the
embedded objects which give rise to the emission spectra seen at $K$.

There are several other stars 
with similar or slightly lower brightness indicated in Figure~\ref{cmdsub},
but with apparently normal colors for hot stars. This suggests a mixture of 
massive objects with and without circumstellar (possibly disk)  
signatures as seen in 
M17 \citep{hhc97}. Like M17, the YSOs indicate a very young age for the cluster.

The situation in W42 may be compared to other stellar clusters
in giant H~II regions. For M17, 
\citet{ldmg91} report that the vast majority of cluster stars they studied
have infrared excesses. This result should be confirmed with higher
spatial resolution images; the 
photometry presented by \citet{ldmg91} was performed by summing the flux
in 4.8$''$ diameter apertures on their 0.8$''$ \pix \ images.
\citet{ldmg91} note that
their data set was analyzed with profile fitting 
by \citet{gmft91} who obtain similar results, and \citet{hhc97} clearly
show that at least some of the massive stars in M17 have disk--like 
spectroscopic features.
\citet{gmft91} find far fewer stars with excess emission in the Orion
Trapezium cluster and attribute this to a mixture of
older and younger stars. However, \citet{zmw93} reach a different conclusion
regarding the ages of the Orion cluster stars, and we will discuss this 
further in the next section.
In W43 (Paper I), only a handful of objects appear in the excess emission 
region of
the color--color plot and none of the three spectroscopically classified 
hot stars do. The brightest object in W43 
exhibits Wolf--Rayet features in its $K-$band spectrum suggesting an older
age relative to W42. 
This is similar
to NGC~3603, R136, and the Arches 
(see the discussion in Paper I and references therein), and the
implication is that while such stars may still be core hydrogen burning, 
they are not on the ZAMS.

These comparisons need to account for the fact that
the extinction is generally greater in W42 and (much so) in W43, so that many 
objects are not detected at $J$ or $H$ and hence do not 
appear in the color--color plot. 
Clearly, high spatial resolution, homogeneous data sets, each 
analyzed in detail for for completeness
at $J$, $H$, and $K$ would be useful in assessing the intrinsic 
fractions of stars
with excess emission in embedded clusters in giant H~II regions.

Without spectra, it is not possible to identify the nature of the remaining
objects in the color excess region of Figure~\ref{cc}, though in the 
context of this young cluster, it is likely
that some are PMS stars or protostars. 
It is also clear from Figure~\ref{cc} that some YSOs can occupy the 
same region of color space as normal stars. 
Some of the cluster stars in Figure~\ref{cc}
near ($H-K$ , $J-H$ $=$ 1,2) where the dispersion in color is larger
might be a combination of 
WTTS with essentially normal stellar colors and 
CTTS/Herbig AeBe stars with smaller excess emission, the actual
nature depending also on the luminosity of any particular object. 
Alternately, these could
be normal main sequence stars.
Finally, some 
could be normal K or M giants in the background, though
it is unlikely that many are, given the large number of stars in close 
association with the central region of the cluster and the fact that we have
subtracted a background component from the CMD.
 
Inspection of Figures~\ref{cmd} and \ref{cc} indicates that W42~\#3 has varied 
in brightness
and color between the 1998 and 1999 observations presented here, becoming
fainter at $K$ and redder. Color and brightness variations in YSOs can be
explained by a combination of changes in the accretion luminosity produced
by the circumstellar disk and changes in the extinction toward the source.
\citet{smwh96} monitored a sample of YSOs for color changes, finding 
the slopes of the colors (measured over day to year timescales) 
in the $J-H$ vs. $H-K$ color--color diagram were intermediate
between the CTTS locus (excess color arising from disk luminosity) and pure
variations along the interstellar reddening lines such as might
occur when clumps of obscuring material pass in front of the 
line of sight to the YSO. \citet{smwh96} found
some YSO variations were consistent with one or the other effect dominating.
The color variation for W42~\#3 appears most consistent with a change in
the local extinction.

\subsection{The $K-$band Luminosity Function}

The KLF has been used as the basis for mass function determinations in young
embedded or obscured massive star clusters. \citet{ldmg91} found that the 
M17 cluster KLF had a slope consistent with the \citet{s55} initial 
mass function (IMF) value if
the M17 stars were normal main sequence type, but noting that this was 
difficult to reconcile with their finding that essentially all the M17 cluster
stars had infrared excesses. Only in the case where such excesses arise in 
passive disks could their mass function determination still follow from the
KLF. \citet{gmft91} present a KLF for the Trapezium cluster in Orion (M42).
They found the slope of the KLF to be inconsistent with that
for the \citet{s55} initial mass function (IMF), 
hypothesizing that there are substantial numbers of older stars in M42 as 
well as very young ones. However, \citet{zmw93} found that the KLF in M42
was consistent with a population of only very young stars if
PMS evolutionary tracks were used instead of assuming the cluster stars
were on the main sequence. \citet{zmw93} show that
deuterium burning on the PMS
can cause peaks in the KLF which are a function of age thus producing 
luminosity functions different from expected if main sequence mass--luminosity
relations are assumed. More recently, 
the detailed optical imaging/spectroscopic investigation
of \citet{h97} exquisitely details the young main sequence and
PMS stellar population in the Orion Trapezium cluster, clearly demonstrating
the overall youth of the cluster.

\citet{feal99} presented a mass function determination for the Arches cluster
located near the Galactic center (GC). They find that the Arches
cluster may have upwards of 100 O stars and an age of $\sim$ 2 Myr. Using their
observed KLF and a mean value for $A_K$, these authors determined the mass
function by relating the $K$ magnitudes to stellar mass using 
the \citet{mmssc94} stellar evolutionary models. \citet{feal99} find a 
mass function in the Arches which is significantly flatter than \citet{s55}.
They attribute this result to the different pre--conditions for star formation
in the central few hundred parsecs of the Galaxy which they argue should favor
higher mass stars. This is in contrast to the situation elsewhere in the 
Galaxy and in the Large Magellanic Cloud (LMC) \citep{mh98}. \citet{mh98} find
from a combination of spectroscopy and imaging that 
the IMF is similar for OB associations and dense massive star clusters
(most notably R136 in the LMC) and in agreement with a \citet{s55}
IMF.

The preceding discussion indicates that care must be taken in transforming the
KLF into a mass function, including the effects of PMS evolution. In this
sense, the determination of a mass function for W42 is premature given the 
implied young age and lack of transformations between PMS models and
near infrared colors (which should include the effects of associated disk
emission). However, a comparison to the different KLFs is warranted. 

In \S3.1 we derived the KLF for the inner 30$''$ central cluster in W42
and showed that it
should be complete to \apge 80 $\%$ for $K \leq$ 15 mag. The measured slope
of the cumulative counts in the central 30$''$ KLF ($K \leq$ 15, correcting
for incompleteness) 
is 0.38 $\pm$ 0.016. 
If we consider only stars in the background subtracted CMD 
(Figure~\ref{cmdsub}, the measured slope ($K \leq$ 15) is 0.36 $\pm$ 0.011. 
The slope in our background annulus
(30$''$ to 50$''$, $K \leq$ 15, correcting for incompleteness) is 
0.51 $\pm$ 0.017. These values may be compared to the 
results of \citet{ldmg91}, \citet{zmw93}, and \citet{feal99}.
For convenience , we compare the power--law slope to the {\it cumulative}
KLF which \citet{ldmg91} also calculated for their data.
In M17, \citet{ldmg91} report a slope of 0.26, which they claim is 
consistent with the \citet{s55} IMF. We have calculated approximate slopes
for \citet{zmw93} and \citet{feal99} from figures of the
published luminosity functions; 
these are 0.39 $\pm$ 0.016 and 0.28 $\pm$ 0.013, respectively. 
\citet{feal99} conducted completeness
detailed tests and corrected the published counts. \citet{ldmg91} conducted more
rudimentary checks on completeness and report the KLF for the magnitude range
they believe to be complete. \citet{zmw93} and \citet{gmft91} conducted no 
completeness tests using artificial stars, to the best of our knowledge, though
\citet{zmw93} claim the KLF is complete. All the KLFs discussed here are defined
for the central clusters in the respective star forming regions, and none
have been corrected for reddening. Thus our comparison is only for the slopes
of the KLFs and does not account for possible differences resulting from
differential reddening.

The KLF in W42 appears to be more similar
to that in Orion (i.e. relatively steep) than in M17 or the
Arches. Given the likely presence of YSOs as discussed 
above, we may be seeing the effects of PMS
evolution on the number counts as is the case for the Trapezium. 
Both the Arches and M17 clusters clearly have
more massive stars than are present in the central cluster of W42 or the 
Trapezium \citep{ceal96, hhc97}. The apparent KLFs in these latter 
two clusters appear
at least superficially similar, though \citet{feal99} and \citet{ldmg91} reach 
rather different conclusions when converting to mass functions. \citet{feal99}
derive an IMF which is much flatter ($\sim$ factor of 2) 
than \citet{s55}, while \citet{ldmg91}
find a mass function which is possibly consistent with \citet{s55}.

\subsection{Distance}

In \S3.2 we classified the spectrum of the bright central star, W42~\#1,
as an early to mid O type ($\approx$ O5--O6). Several lines of evidence 
presented above suggest that the W42 cluster is quite young. 
If we take W42~\#1 to be 
on the ZAMS then its apparent brightness would give a distance to the 
cluster of 2.2 kpc, considerably closer than the radio distance (3.7 kpc, see
\S3). In \S3.2, we argued that the spectrum of W42~\#1 was most
similar to those for the dwarf or giant luminosity classes.
Using the average $M_V$ from \citet{vgs96} for dwarf stars 
and colors from \S3, gives
2600$_{700}^{1000}$ kpc. If the giant star $M_V$ from \citet{vgs96} 
is used, then 
the distance estimate becomes 3400$_{900}^{1200}$ pc. The uncertainty in the
distance estimate is completely dominated by the 
luminosity class assumed and the scatter in the intrinsic $M_V$ of
O stars ($\pm$ 0.67 mag). The uncertainties in the reddening and apparent
magnitude are negligible in comparison: a few percent for $m_K$ and \aple 
10 $\%$ for $A_K$, where the largest part of the uncertainty is in the 
choice of reddening law (see Mathis 1990). These distance estimates are in
agreement with \citet{ldw85} who argued that W42 (G25.4SE) was at the 
near distance given by the radio recombination line velocity and 
Galactic rotation model. Depending on the true luminosity of W42~\#1, the 
cluster may even be somewhat closer. 
\citet{sbm78} estimated the Lyc luminosity of W42 to be 
8.2$\times~10^{50}~s^{-1}$
assuming the far kinematic distance (13.4 kpc). 
Adopting the ZAMS distance (2.2 kpc) as indicated by the young nature of the
cluster (stars W42~\#2 and \#3)
considerably 
reduces the expected ionizing flux to 2$\times10^{49} s^{-1}$.

In Figure~\ref{cmdsub}, we have plotted a zero age main sequence (ZAMS) derived
from the models of \citet{ssmm92} (Z $=$ 0.02); see \S3.1.3. 
We have placed the ZAMS in Figure~\ref{cmdsub}
at 2200 pc, assuming that
the W42~\#1 is on the ZAMS. 
In this case, the (incomplete) 
cluster sequence reaches early K type stars. For the Trapezium \citep{zmw93}
claim that the KLF shows an intrinsic peak well above the sensitivity of their
$K-$band images. This peak is due to the PMS evolution of the lower mass
stars and is sensitive to age on the PMS. For the close distance to W42
implied if W42~\#1 is on the ZAMS, the similarity found above for the
KLF of W42 compared to the Trapezium then suggests a similar effect on the
KLF may be at work. I.e., the fainter magnitudes may correspond to
lower mass PMS stars. In the present case, we are not claiming to see a
real turn over in the KLF (as Zinnecker et al. do in Orion) 
since this occurs where the $K-$band counts are demonstrably incomplete due
to crowding.

For a distance of 2.2 kpc, the cluster has a radius of
0.32 pc at 30$''$. The surface density in the inner 10$''$ (Figure~\ref{surf})
is then within a factor of two for that in the Trapezium \citep[ $\sim$ 6500
pc$^{-2}$]{ms94} not accounting for the substaintial incompleteness in the 
present case. W42 is thus quite dense, and this is another indication of 
its young age \citep{ms94, feal99}.

The ZAMS clearly demonstrates the lack of any sensitivity to temperature in
the hot stars at near infrared wavelengths. This means that traditional
methods of cluster fitting (in the observational color--magnitude
plane) for age or distance can not be used and only
through suitable numbers of infrared spectra (resulting in \mbol \ 
and \teff)
will the cluster 
properties truly emerge. 
The lack of temperature sensitivity in the near infrared colors
does, however, lead to accurate estimates of the 
foreground extinction.

\section{SUMMARY}

We have presented high spatial resolution $J$, $H$, and $K$ images of the
massive star cluster at the heart of the giant H~II region W42. Our $K-$band
spectra of three of the brightest four stars in the central 30$''$ of the 
cluster indicate a very young population. The brightest star W42~\#1 is 
classified as kO5--O6 based on the system of \citet{hhc97}. Such stars
are typically associated with MK O5--O6.5 dwarfs. W42~\#2 and W42~\#3 show
no stellar absorption features. This fact, combined with
their position
in the excess emission band of the 
$J-H$ vs. $H-K$ color--color plot, leads us to classify them as YSOs.

The KLF was computed and compared to other massive star clusters. The KLF 
in W42 appears more similar to that of the Trapezium than to the more
massive clusters in M17 and the Arches. The steepness of the KLF in the 
Trapezium has been attributed to the PMS stars present there, and it is
possible that a similar effect is present in W42 given
our spectroscopic and imaging evidence for YSOs.

Our spectrum of  W42~\#1 confirms the results of \citet{ldw85} that W42 
(G25.4SE) can not be at the far kinematic distance. Earlier estimates of the 
Lyc output from this giant H~II region must therefore be revised downward
(by $\sim$ an order of magnitude). If W42~\#1 is on the ZAMS, 
as we argue based on the presence of young stellar objects, then the 
cluster is even closer than the near kinematic distance (3.7 kpc). We
estimate 2.2 kpc. In this case, W42 should not be considered a giant H~II region
as defined in \S1.
Our spectrum of W42~\#1 is not sensitive to luminosity class, though 
it suggests a dwarf or giant classification. 

PSC appreciates continuing support from the National 
Science Foundation. We wish to acknowledge the 
continuing excellent support received from the CTIO mountain staff, 
particularly from night assistants Hern\'{a}n Tirado, Patricio Ugarte, and 
Alberto Zu\~{n}iga.

%REFERENCES

\newpage

%FIGURES

\newpage

%TABLES

%\pagestyle{empty}
\begin{deluxetable}{lrrrrrrr}
\tablecaption{ZAMS Properties\label{tabzams}}
\tablewidth{0pt}
\tablehead{
\colhead{SpType}& 
\colhead{\teff\tablenotemark{a} }& 
\colhead{Mass \tablenotemark{b} }& 
\colhead{\mbol\tablenotemark{b} }& 
\colhead{$M_V$\tablenotemark{c} }& 
\colhead{$M_K$ }& 
\colhead{$V-K$\tablenotemark{d} }& 
\colhead{$H-K$\tablenotemark{d} } }
\startdata
O3 & 51230 & 89.7 & $-$10.28 & $-$5.73 & $-$4.80 & $-$0.93 & $-$0.05\\
O4 & 48670 & 65.1 & $-$9.74 & $-$5.34 & $-$4.41 & $-$0.93 & $-$0.05\\
O4.5 & 47400 & 56.4 & $-$9.50 & $-$5.17 & $-$4.24 & $-$0.93 & $-$0.05\\
O5 & 46120 & 49.3 & $-$9.24 & $-$5.00 & $-$4.07 & $-$0.93 & $-$0.05\\
O5.5 & 44840 & 43.6 & $-$8.97 & $-$4.82 & $-$3.89 & $-$0.93 & $-$0.05\\
O6 & 43560 & 38.9 & $-$8.70 & $-$4.62 & $-$3.69 & $-$0.93 & $-$0.05\\
O6.5 & 42280 & 34.9 & $-$8.41 & $-$4.43 & $-$3.50 & $-$0.93 & $-$0.05\\
O7 & 41010 & 31.5 & $-$8.12 & $-$4.23 & $-$3.30 & $-$0.93 & $-$0.05\\
O7.5 & 39730 & 28.6 & $-$7.84 & $-$4.04 & $-$3.11 & $-$0.93 & $-$0.05\\
O8 & 38450 & 26.0 & $-$7.55 & $-$3.85 & $-$2.92 & $-$0.93 & $-$0.05\\
O8.5 & 37170 & 23.7 & $-$7.28 & $-$3.68 & $-$2.75 & $-$0.93 & $-$0.05\\
O9 & 35900 & 21.6 & $-$7.01 & $-$3.51 & $-$2.62 & $-$0.89 & $-$0.05\\
O9.5 & 34620 & 19.7 & $-$6.75 & $-$3.36 & $-$2.49 & $-$0.87 & $-$0.05\\
B0 & 33340 & 17.9 & $-$6.49 & $-$3.21 & $-$2.36 & $-$0.85 & $-$0.05\\
B0.5 & 32060 & 16.3 & $-$6.23 & $-$3.07 & $-$2.28 & $-$0.79 & $-$0.04\\
B1 & 21500 & 7.2 & $-$3.52 & $-$1.70 & $-$0.94 & $-$0.76 & $-$0.04\\
B2 & 18000 & 5.4 & $-$2.31 & $-$0.92 & $-$0.25 & $-$0.67 & $-$0.04\\
B3 & 15500 & 4.3 & $-$1.39 & $-$0.36 & 0.21 & $-$0.57 & $-$0.03\\
B5 & 13800 & 3.6 & $-$0.74 & 0.09 & 0.52 & $-$0.43 & $-$0.02\\
B7 & 12200 & 3.0 & $-$0.07 & 0.59 & 0.89 & $-$0.30 & $-$0.02\\
B9 & 10600 & 2.5 & 0.70 & 1.15 & 1.29 & $-$0.14 & $-$0.01\\
A0 & 9850 & 2.2 & 1.13 & 1.44 & 1.44 & 0.00 & 0.00\\
A2 & 9120 & 2.0 & 1.59 & 1.76 & 1.63 & 0.13 & 0.01\\
A5 & 8260 & 1.8 & 2.23 & 2.15 & 1.80 & 0.35 & 0.02\\
A7 & 7880 & 1.7 & 2.55 & 2.42 & 1.97 & 0.45 & 0.02\\
F0 & 7030 & 1.4 & 3.36 & 3.23 & 2.44 & 0.79 & 0.04\\
F2 & 6700 & 1.3 & 3.72 & 3.61 & 2.68 & 0.93 & 0.05\\
F5 & 6400 & 1.2 & 4.07 & 3.98 & 2.97 & 1.01 & 0.06\\
F8 & 6000 & 1.1 & 4.58 & 4.52 & 3.40 & 1.12 & 0.06\\
G0 & 5900 & 1.1 & 4.71 & 4.67 & 3.45 & 1.22 & 0.07\\
G2 & 5770 & 1.1 & 4.89 & 4.86 & 3.46 & 1.40 & 0.08\\
G5 & 5660 & 1.0 & 5.05 & 5.04 & 3.49 & 1.55 & 0.08\\
G8 & 5440 & 1.0 & 5.37 & 5.41 & 3.81 & 1.60 & 0.09\\
K0 & 5240 & 0.9 & 5.69 & 5.76 & 4.01 & 1.75 & 0.10\\
K2 & 4960 & 0.8 & 6.15 & 6.30 & 4.05 & 2.25 & 0.13\\
\enddata
\tablenotetext{a}{\teff \ vs. SpType from Vacca et al. (1996) and Johnson (1966)}
\tablenotetext{b}{Determined from \teff \ and models of Schaller et al. (1992)}
\tablenotetext{c}{Determined from \teff \ and bolometric correction 
relation given by Vacca et al. (1996, \teff \ $>$ 28000) and Malagnini et al. 
(1986, \teff \ $\leq$ 28000)}
\tablenotetext{d}{Koorneef (1983), corrected to the CIT/CTIO system; see text} 
\end{deluxetable}

\begin{thebibliography}{dummy}

\bibitem[Blum et al.(1997)]{brcfs97} Blum, R.D., Ramond, T.M., Conti, P.S., 
Figer, D.F, \& Sellgren, K. 1997, AJ, 113, 1855

\bibitem[Blum et al.(1999)]{bdc99}  Blum, R.D., Damineli, A., \& Conti, P.S.
1999, \aj, 117, 1392 (Paper I)

\bibitem[Brander et al.(1999)]{beal99} Brander, W., et al. 1999, \baas, 194.6808B

\bibitem[Carter(1990)]{c90} Carter, B. S. 1990, \mnras, 242, 1

%\bibitem[Carter \& Meadows(1990)]{cm95} Carter, B. S., \& Meadows V. S. 
%1995, \mnras, 276, 734

%\bibitem[Conti et al.(1995)]{cea95} Conti, P. S., Hanson, M. M., Morris, 
%P. W., Willis, A.  J., \& Fossey, S. J. 1995, \apj, 445, L35

\bibitem[Cotera et al.(1996)]{ceal96} Cotera, A., Erickson, E. F., Colgan, S. W. J., Simpson, J. P., Allen, D. A., \& Burton M. 1996, \apj, 461, 750

%\bibitem[Cotera \& Simpson(1997)]{cs98} Cotera, A. \& Simpson J. 1997, 
%AAS meeting 191, abstract 114.03

\bibitem[DePoy et al.(1993)]{dabfo93}DePoy, D. L., Atwood, B., Byard, P., 
Frogel, J. A., \& O'Brien, T. 1993, in SPIE 1946, ``Infrared Detectors and 
Instrumentation,'' pg 667 

%\bibitem[Drissen et al.(1995)]{dmws95} Drissen, L., Moffat, A. F. J., 
%Walborn, N. R., \& Shara, M. M. 1995, \aj, 110, 2235

%\bibitem[Eenens et al.(1991)]{eww91} Eenens, P. R. J., Williams, P. M., \& 
%Wade, R. 1991, \mnras, 252, 300

\bibitem[Elias et al.(1982)]{efmn82} Elias, J. H., Frogel, J. A., Matthews, K., \& Neugebauer, G.  1982, \aj, 87, 1029

\bibitem[Figer et al.(1997)]{fmn97} Figer, D. F., McLean, I. S., \& Najarro, F. 1997, \apj, 486, 420

%\bibitem[Figer et al.(1998)]{fea98} Figer, D. F., Najarro, F., Morris, M., 
%McLean, I. S., Geballe, T. R., Ghez, A. M., \& Langer, N. 1998, \apj, 506, 384 

\bibitem[Figer et al.(1999a)]{fmm99} Figer, D. F., McLean, I. S., \& Morris, M.
1999, \apj, 506, 384 

\bibitem[Figer et al.(1999b)]{feal99} Figer, D. F., Kim, S. S., Morris, M.,
Serabyn, E., Rich, R. M., McLean, I. S., 1999, \apj, in press

\bibitem[Frogel et al.(1978)]{fpam78} Frogel, J. A., Persson, S. E., 
Matthews, K., \& Aaronson, M. 1978, \apj, 220, 75

\bibitem[Gatley, et al.(1991)]{gmft91} Gatley, I., Merrill, K. M., 
Fowler, A. M., \& Tamura, M. 1991, in Astrophysics with Infrared Arrays, ed. R. Elston, ASP Conf. Ser. 14, p230

\bibitem[Hanson et al.(1996)]{hcr96} Hanson, M. M., Conti, P. S., \& Rieke, M. J. 1996, \apjs, 107 281
 
\bibitem[Hanson et al.(1997)]{hhc97} Hanson, M. M., Howarth, I.D., \& Conti, P.S. 1997, ApJ, 489, 698

\bibitem[Hanson et al.(1998)]{hrl98} Hanson, M. M., Rieke, G., \& Luhman K. L.
1998, \aj, 116 1915

\bibitem[Hartmann et al.(1993)]{hkc93} Hartmann, L., Kenyon, S. J., \& 
Cavalet, N. 1993, \apj, 407, 219

\bibitem[Hester et al.(1996)]{heal96} Hester, J. J., et al. 1996, \aj, 111, 2349

\bibitem[Hillenbrand(1997)] {h97} Hillenbrand, L. A. 1997, \aj, 113, 1733

\bibitem[Johnson(1966)]{j66} Johnson, H. 1966, \araa, 4, 193

\bibitem[Koorneef(1983b)]{k83.1} Koornneef, J. 1983, \aaps, 51, 489

\bibitem[Koorneef(1983a)]{k83} Koornneef, J. 1983, \aap, 128, 84

\bibitem[Lada et al.(1991)]{ldmg91} Lada, C. J., DePoy, D. L., Merrill, K. M.,
\& Gatley, I. 1991, \apj, 374, 533

\bibitem[Lada \& Adams(1992)] {la92} Lada, C. J. \& Adams F. C. 1992, \apj, 393, 278

\bibitem[Lester et al.(1985)]{ldw85} Lester, D.F., Dinerstein, H.L., Werner, M.W., Harvey, P.M., Evans II, N.J., \& Brown, R.L. 1985, \apj, 296, 565

\bibitem[Malagnini et al.(1986)]{mmrk86} Malagnini, M. L., Morossi, C., Rossi, L., \& Kurucz, R. L. 1986, \aap, 162, 140

\bibitem[Massey et  al.(1989)]{mpg89} Massey, P., Parker, J., \& Garmany, C. D.
1989, \aj, 98, 1305

\bibitem[Massey et  al.(1995)]{mjd95} Massey, P., Johnson, K., \& DeGioia-Eastwood, K. 1995, ApJ, 454, 151

\bibitem[Massey \& Hunter(1998)]{mh98}  Massey, P. \& Hunter, D. A. 1998, \apj, 493, 180

\bibitem[Mathis(1990)]{m90} Mathis, J.S. 1990, \araa, 28, 37

\bibitem[McCaughrean \& Stauffer (1994)]{ms94} McCaughrean, M. J. \& Stauffer, 
J. R. 1994, \aj, 108, 1382

\bibitem[Meyer et al.(1997)]{mch97} Meyer, M. R., Calvet, N., \& Hillenbrand, L. A. 1997, \aj, 114, 288

\bibitem[Meynet et al.(1994)]{mmssc94}Meynet G., Maeder A., Schaller G., Schaerer D., Charbonnel C. 1994, \aap, 103 97

\bibitem[Morris \& Serabyn(1996)]{ms96} Morris, M. \& Serabyn, E. 1996, \araa, 34, 645

\bibitem[Olivia \& Origlia(1992)]{oo92} Olivia, E. \& Origlia, L. 1992, \aap, 254, 466

\bibitem[P\'{e}rez \& Elston(1998)]{pe98} P\'{e}rez, G. \& Elston, R. 1998, Proc. SPIE, 3352, 328

\bibitem[Persson et al.(1998)]{peal98} Persson, S. E., Murphy, D. C., Krzeminski, W., \& Roth, M. 1998, \aj, 116, 2475

\bibitem[Pezzuto et al.(1997)]{psl97} Pezzuto, S., Strafella, F., 
Lorenzetti, D. 1997, \apj, 485, 290

\bibitem[Salpeter(1955)]{s55} Salpeter, E. E. 1955, \apj, 121, 161

\bibitem[Schaller et al.(1992)]{ssmm92} Schaller G., Schaerer D., Meynet G., Maeder A., 1992, \aaps, 96, 269

\bibitem[Schecter et al.(1993)]{sms93} Schecter, P. L., Mateo, M. L., \& Saha, A. 1993, \pasp, 105, 1342

%\bibitem[Schmidt--Kaler(1982)]{sk82} Schmidt--Kaler, T. 1982, in 
%Landolt--B\"{o}rstein, New Series, group VI, vol. 2, ed. K. Schaifers 
%\& H. H. Voigt (Berlin: Springer--Verlag), 1

\bibitem[Skrutskie et al.(1996)]{smwh96} Skrutskie, M. F., Meyer, M. R., 
Whalen, D., \& Hamilton C. 1996, \aj, 112, 2168

\bibitem[Smith et al.(1978)]{sbm78} Smith, L.F., Biermann, P. \& Mezger, P.G. 1978, \aap, 66, 65

\bibitem[Vacca et al.(1996)]{vgs96} Vacca, W. D., Garmany, C. D., \& Shull, J. M. 1996, \apj, 460, 914

\bibitem[Zinnecker et al.(1993)]{zmw93} Zinnecker, H., McCaughrean, M. J.,
\& Wilking, B. A. 1993, in Protostars and Planets III, eds E. Levy \& J. Lunine,
(Tucson: University of Arizona Press), p429

\end{thebibliography}
\end{document}